\documentclass[journal=apchd5,manuscript=letter,layout=twocolumn]{achemso}

\usepackage[version=3]{mhchem} 
\usepackage{multicol}
\usepackage{soul}
\usepackage{xcolor}

\author{Ankit Dulat}
\email{dulatankit@gmail.com}
\author{Sk Rakeeb}
\author{Sagar Dam}
\author{Amit D. Lad}
\author{Yash M. Ved}
\affiliation{Tata Institute of Fundamental Research, 1 Homi Bhabha Road, Colaba, Mumbai 400 005, India}
\author{Sergey Kruk}
\affiliation{Australian Research Council Centre of Excellence QUBIC, IBMD, School of Mathematical and Physical Sciences, University of Technology Sydney, Australia}
\author{G. Ravindra Kumar}
\email{grk@tifr.res.in}
\affiliation{Tata Institute of Fundamental Research, 1 Homi Bhabha Road, Colaba, Mumbai 400 005, India}

\title{Coherent Control of Relativistic Electron Dynamics in Plasma Nanophotonics}

\begin{document}







\twocolumn[
\begin{@twocolumnfalse} 
\maketitle
\begin{abstract}
Intense femtosecond laser pulses interacting with solids can drive electrons to relativistic energies, enabling miniaturized particle accelerators and bright extreme-ultraviolet light sources. \textit{In-situ} space-time control of these electrons is crucial for developing next-generation laser-based accelerators but remains extremely challenging. We present a novel approach to achieve such control by manipulating the local fields driving these electrons using a nanoengineered dielectric nanopillar target. We demonstrate via experiments and simulations that this sub-femtosecond and nanometer-scale control enables enhanced electron acceleration and control of the directionality of relativistic electrons over a wide angular range and predicts the coherent formation of sub-femtosecond electron bunches from the nanopillars. This research bridges nanophotonics and strong-field plasma physics, offering new opportunities for \textit{in-situ} control of high-energy particles and advancements in plasma technology.
\end{abstract}


\end{@twocolumnfalse}]


\section{Introduction}
Recent years have witnessed ever growing interest in plasma nanophotonics \cite{rocca2024,Ciappina2017} -- a field studying the interaction of ultrahigh-power lasers with matter structured at the nanoscale. These extreme interactions have facilitated the creation of ultrahigh-energy-density states \cite{Hollinger2020,Purvis2013,Bargsten2017,Kaymak2016}, exceptionally bright and highly energetic particles \cite{Moreau2020,Jiang2016,Ceccotti2013}, tabletop-scale fusion \cite{Curtis2021,Curtis2018}, relativistic attosecond electron pulses \cite{Cardenas2019,Elkind2023}, and extreme ultraviolet (XUV) light sources \cite{Hollinger2017,Rajeev2003,Mondal2011,Bagchi2011}. Relativistic electrons generated from these interactions \cite{Gourab2012,Moreau2020,Jiang2016} are the key to generating extreme states and are crucial for applications. Spatio-temporal control over relativistic electrons is highly desirable yet extremely difficult to achieve. It requires precise manipulation of local fields on sub-femtosecond (fs) time scales simultaneously with nanometric spatial scales, as these fields strongly influence the trajectories of the electrons \cite{Li2006}. Previous studies with flat \cite{Li2006,Cho2009,Wang2010,Mao2012} and nanostructured targets \cite{Cristoforetti2017,Habara2021,Tian2014,cristoforetti2014} have shown the critical dependence of the electron emission on the laser absorption mechanism \cite{Cho2009,Wang2010} and pre-plasma scale length \cite{Mao2012,Li2006}. However, in high-intensity laser-plasma experiments, multiple absorption mechanisms may operate simultaneously, contingent upon the laser and target parameters, and nanometer-level control of pre-plasma remains extremely challenging. These challenges have hindered efforts to control relativistic electron dynamics.

Studies of structured targets have so far been mainly limited to the enhancements of electromagnetic and ion emissions \cite{Moreau2020,Jiang2016,Ceccotti2013,Curtis2021,Hollinger2017,Rajeev2003,Mondal2011,Bagchi2011,Gourab2012, cristoforetti2014}. However control over the larger parameter space of the emission of electromagnetic radiation, ions, and electrons remains largely unexplored. In particular, opportunities for \textit{in-situ} control of the directionality of the electron beams via target nanostructuring have not been harnessed.

At present, control over the emitted relativistic electrons is performed predominantly with post-interaction approaches (upon which electron trajectories are controlled after the generation): either using external electromagnets (magnetic collimation) \cite{Bailly2018,Kar2016} or specially designed targets \cite{Kodama2004} with resistivity gradients (resistive collimation) \cite{Ramakrishna2010,Robinson2007,Bell2003}. However, these approaches offer only limited control, primarily affecting the electron beam divergence.  A much more attractive approach relies on \textit{in-situ} manipulation \cite{Yeung2017,Borot2012}, controlling local fields (both laser and plasma fields) during the interaction between the fs pulse and the plasma. This allows the guiding of the electron beams directly during their acceleration phase.

In underdense gaseous plasmas, particularly in laser wakefield acceleration (LWFA) \cite{Huijts2022,Zhu2020,Vieira2018,He2015,Popp2010,Nerush2009}, \textit{in-situ} control over the directionality of relativistic electron beams has been demonstrated by engineering incident laser pulses, including manipulating the carrier-envelope phase (CEP) and pulse front tilt (PFT) of the driving few-cycle laser pulse. Such waveform-dependent control of electron beams is however only possible with few-cycle laser pulses ($<$ 10 fs), and it is particularly challenging in laser-solid interactions. Indeed, in solids, the laser pulse interacts with the critical surface of the plasma, where the carrier-envelope phase (CEP) is fixed, while in underdense plasmas, the laser pulse can pass through the plasma (as in LWFA), and the CEP can play a vital role \cite{Huijts2022}.

\begin{figure}[!ht]
\centering
\includegraphics[width=1\linewidth]{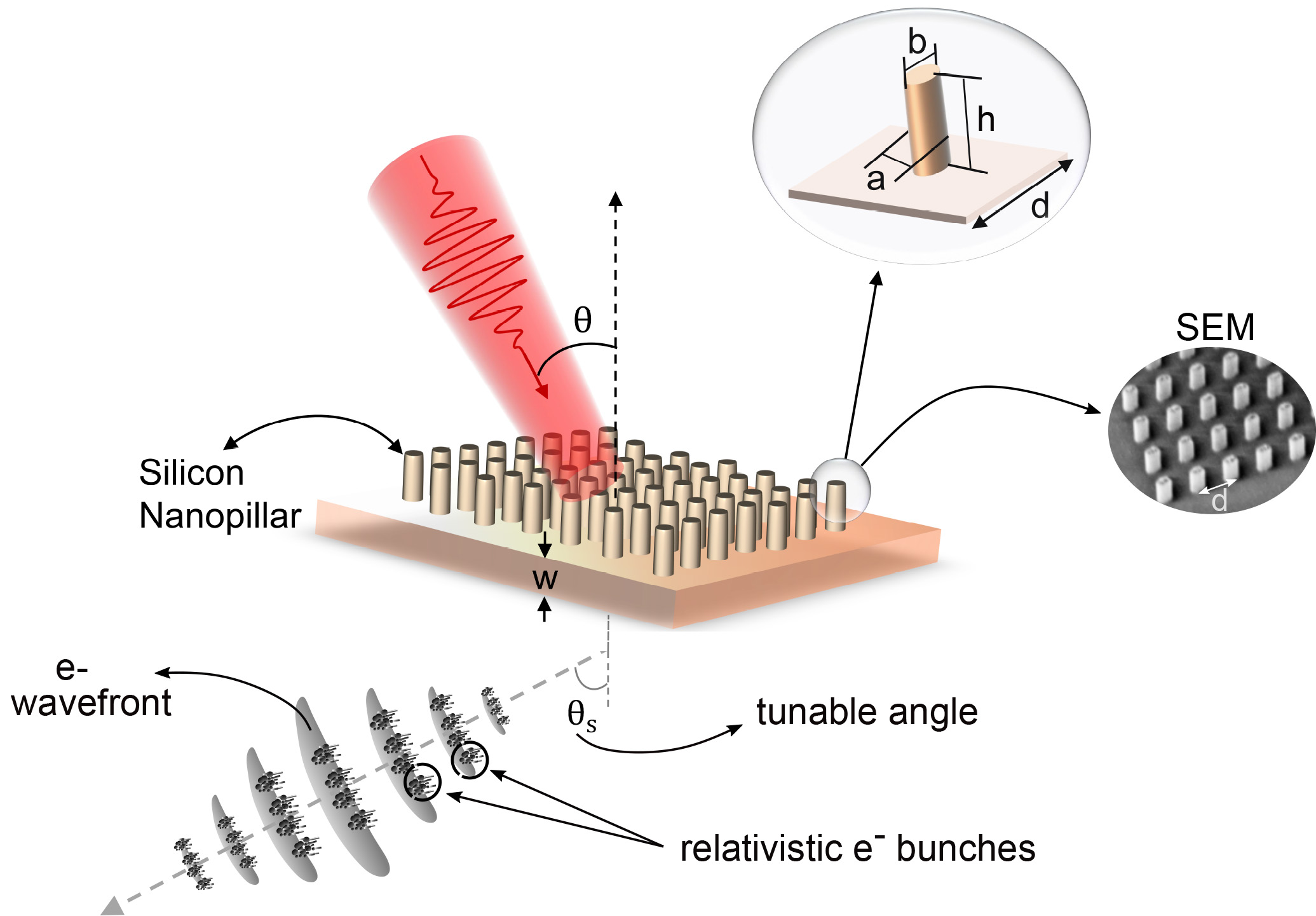}
\caption{Schematic of the interaction of an intense fs laser pulse with a nanostructured target, resulting in control over the directionality and enhanced acceleration of electrons on the rear of the target. The guiding direction of the electron beam is tunable by changing the AOI of the laser pulse. The inset shows the scanning electron microscope (SEM) image of the nanostructured target, with dimensions a = 280 nm, b = 450 nm, h = 700 nm, and d = 720 nm.}\label{fig: Fig1}
\end{figure}
Here, we demonstrate a novel method for \textit{in-situ} spatio-temporal control of relativistic electrons from laser-solid interactions based on structuring solids on the nanoscale. Our experimental measurements and particle-in-cell (PIC)  simulations show both enhanced acceleration and directional control of relativistic electrons in a desired direction. 
In our research, we used a periodic arrangement of vertically aligned, near-wavelength-sized nanopillars as the target. The geometry of the nanopillars allows the manipulation of the local near-fields on the vacuum-plasma interface. We exploit the fact that the near-field pattern is due to the interference of the scattered field (Mie scattering) from neighboring nanostructures of the array, which can be controlled either by changing the geometry and arrangement of the nanostructured elements or by controlling the parameters of the driving laser pulse. We show such control on sub-fs time scales and nanometer spatial scales by changing the angle of incidence (AOI) of the excitation laser pulse.\\

Furthermore, PIC simulations demonstrate that the guided electron beams consist of a train of sub-femtosecond electron pulses. The phase difference of the incident laser between neighboring nanopillars plays a crucial role in controlling the electron dynamics. We further observe that electron bunches emitted from different nanopillars are coherently bunched together, in a light-like wavefront structure. Our approach to controlling electron beams is analogous to the working principle of phased array antennas \cite{Constantine}, where the manipulation of the relative phase of the driver current between neighboring antennas in the array enables the constructive addition of radiation in a desired direction. We also compare our results both experimentally and via PIC simulations with a flat target and show enhancements in the electron flux and the cutoff energy of the electrons for the nanostructured target.

\section*{Results}
\subsection*{A. Experiment}

\begin{figure*}[!ht]
\centering
\includegraphics[width=1\linewidth]{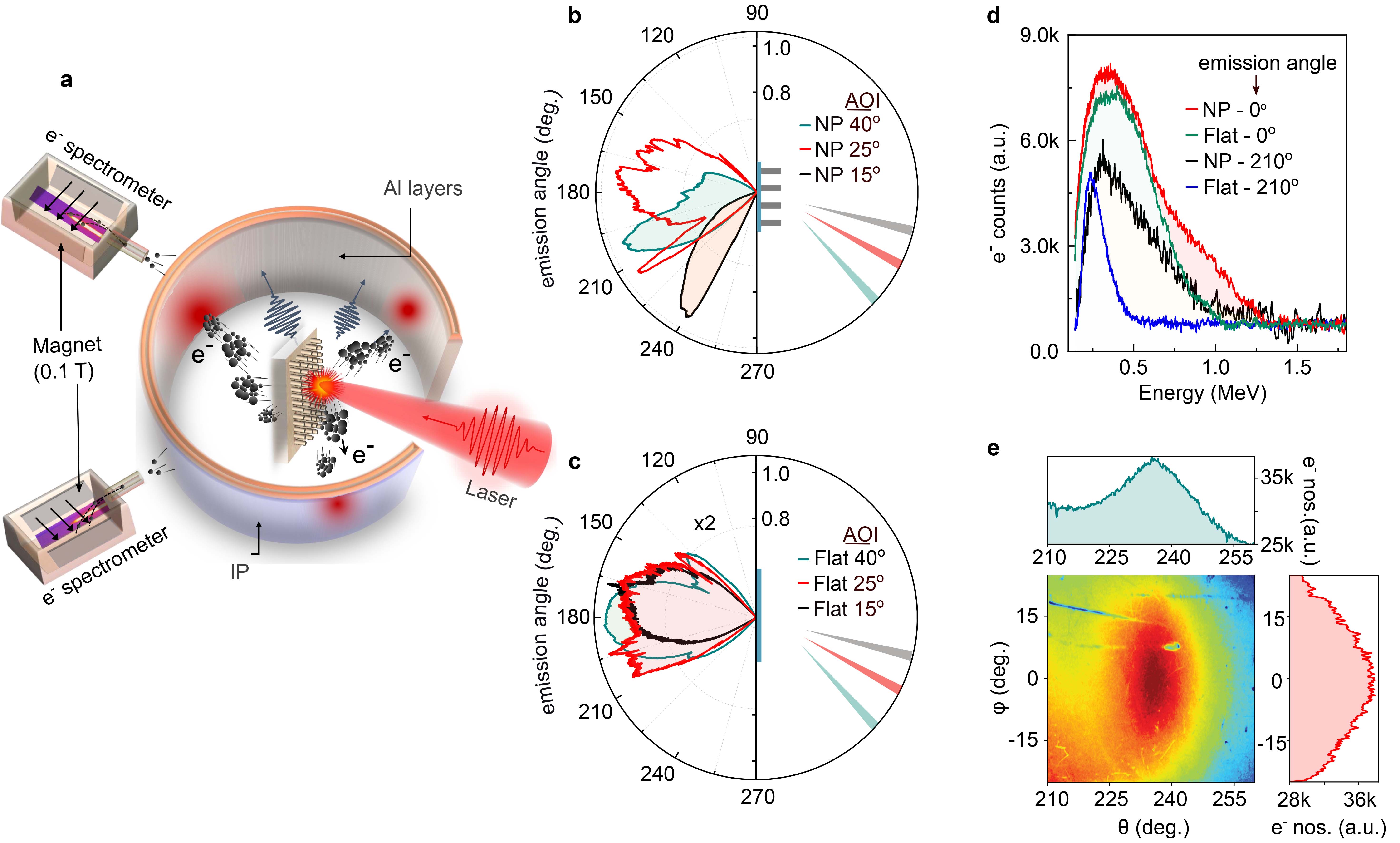}
\caption{\textbf{Experimental results for flat and nanostructured targets.} \textbf{a} Schematic of the experimental setup for measuring the angular distribution of emitted electrons using an imaging plate (IP) and the energy spectrum using magnetic spectrometers. \textbf{b-c} Measured rear-side electron angular distribution with a lower energy cutoff of 100 keV for three different AOIs of $15^{\circ}, 25^{\circ}$, and $40^{\circ}$ for the nanostructured and flat targets, respectively. \textbf{d} Comparison of the electron energy spectrum measured for flat and nanostructured targets along the front normal ($0^{\circ}$) and guiding direction ($210^{\circ}$) for an AOI of $40^{\circ}$. \textbf{e} Measured 2D angular profile of the guided electron beam for an AOI of $15^{\circ}$.}\label{fig: Fig3}
\end{figure*}

The experiments were performed using p-polarized, 25 fs, 800 nm laser pulses focused to a peak intensity of $3\times10^{18}$ W/c$m^2$ on a nanostructured target, as depicted in Fig. \ref{fig: Fig1}. The nanostructured target consists of a periodic arrangement of perfectly aligned elliptically shaped silicon nanopillars, deposited on a 500-$\mu$m-thick quartz substrate. For comparison, we also performed experiments on an equally thick (500 $\mu$m) flat quartz target without nanostructures. See the Methods section for more details on laser parameters and target fabrication. The experimental results for both nanostructured and flat targets are presented in Fig. \ref{fig: Fig3}. Fig. \ref{fig: Fig3}a shows the schematic of the experimental setup. The angular distribution of emitted electrons with energy greater than 100 keV is measured using an imaging plate (IP) detector surrounding the target. The energy spectrum is recorded with magnet-based spectrometers, which can be positioned in the desired orientation, as shown in the schematic. The angular distribution and energy spectrum were measured separately (see Methods for details).

In Fig. \ref{fig: Fig3}b, we show the measured angular distribution of rear-side electrons (energy $>$ 100 keV) from the nanostructured target for three angles of incidence (AOIs) of 15°, 25°, and 40°, respectively. It is very interesting to note that for an AOI of 15° and 40°, a significant fraction of electrons in the rear are emitted at a totally different angle than is usually expected, that is, along the target-normal (180°) and along $\Vec{J}\times\Vec{B}$ ($\sim$140°-165°) \cite{Cho2009}. The emission angle of the electron is closely related to its heating and acceleration mechanisms. Resonance absorption \cite{Freidberg1971} and Brunel heating \cite{F.Brunel1987} are associated with near-normal electron emission, while J$\times$B heating \cite{Kruer1985} is attributed to emission along the laser propagation direction. Emission at a significantly different angle in the nanopillar target indicates a different type of acceleration mechanism, discussed later. A more intriguing observation is the control over the direction of the electron beam by simply varying the AOI of the laser pulse. However, for an AOI of 25°, we observe that the guided electrons are significantly less, and more electrons are still along the target normal and $\Vec{J}\times\Vec{B}$ directions. To investigate this, we repeated the measurement multiple times, which shows that the fraction of guided electrons varies between experiments. This variability is highly dependent on the fabrication quality of the nanostructure and the precise alignment of the laser focal spot on the nanopillars. 
In Figs. S3 and S2 of the Supporting Information (SI), we attach the measured electron angular distribution and the corresponding microscopy images of these laser shots.

For comparison, Fig. \ref{fig: Fig3}c shows the rear-side electron angular distribution on the flat target. In the flat target, the measured rear electron flux is two times less, and most of the electrons are emitted with larger divergence along the target normal and along the $\Vec{J}\times\Vec{B}$ direction, as expected \cite{Cho2009}. The measured angular width of the guided electron beam is 20–35 degrees from the nanopillar target, in contrast to the 60-degree spreading angle from the flat target. Fig. \ref{fig: Fig3}e depicts the measured two-dimensional (2D) angular profile of the guided electron beam measured 15 cm from the target, for an AOI of $15^{\circ}$. As evident, the guided electrons have a beam-like profile. Fig. \ref{fig: Fig3}d shows the comparison of the energy distribution of electrons from nanostructured and flat targets for an AOI of 40°. The energy spectrum was measured along the front-normal direction (0°) and along the direction of the guided electron beam (210°) in the rear of the targets (see Methods sections for details of the measurement). For the structured target, we observe that rear-side electrons have a cut-off energy as high as 1.2 MeV, which is about two and a half times that for flat target (500 keV). It is also interesting to note that the maximum pondermotive energy \cite{Mulser} that can be gained by electrons near the peak of the laser pulse is around 400 keV; however, it is three times lower than the measured cut-off energy of 1.2 MeV for the structured target. To investigate the enhanced energy gain (acceleration mechanism) and control over the directionality of relativistic electrons, we performed particle-in-cell (PIC) simulations, and the results are discussed in the next section.

\begin{figure*}[!ht]
\centering
\includegraphics[width=1\linewidth]{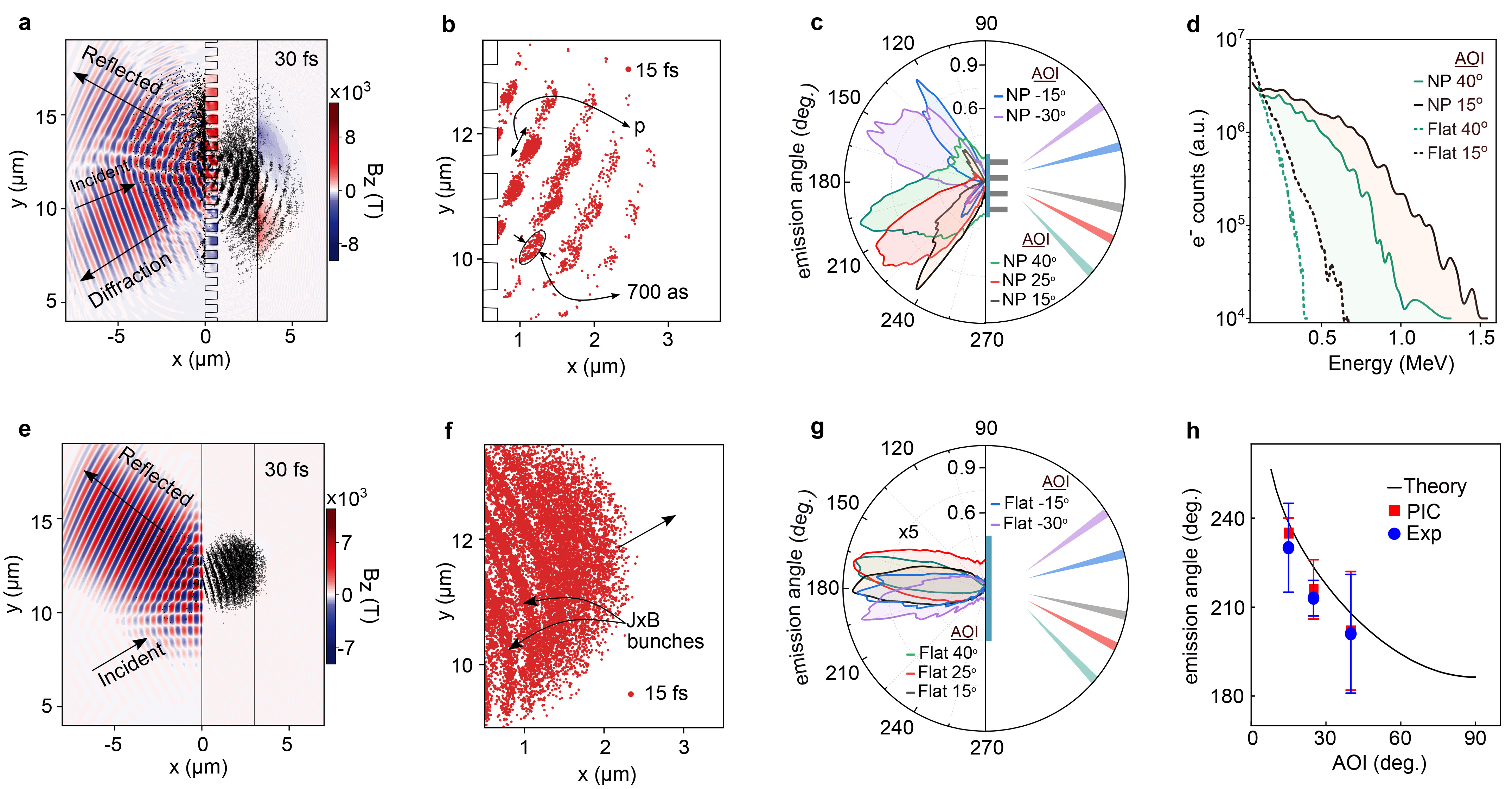}
\caption{\textbf{2D-PIC simulation results for flat and nanostructured targets.} \textbf{a} Magnetic field $B_z$ (color plot) due to the scattered laser field (front) from the nanostructured target and the quasi-static azimuthal magnetic field (rear) due to the current of energetic electrons (shown by black dots). \textbf{b} Sub-fs relativistic electron bunches emitted on the rear of the nanopillars, for an AOI of $40^{\circ}$. \textbf{c} Angular distribution of electrons (energy $>$ 100 keV) for five different AOIs of the laser pulse, as labeled, for the nanostructured target. The colored triangular shapes depict the directions of the incident laser pulse for different AOIs. \textbf{d} Comparison of the electron energy spectrum at the rear of flat and nanostructured targets for two different AOIs of $15^{\circ}$ and $40^{\circ}$. \textbf{e} Magnetic field distribution for the flat target for comparison with \textbf{a}. \textbf{f} Electron distribution on the rear of the flat target, for comparison with \textbf{b}. \textbf{g} Angular distribution of electrons (energy $>$ 100 keV) for five different AOIs of the laser pulse, as labeled, for the flat target. \textbf{h} Comparison of the experimentally measured emission angle of the electron beam with the results of the PIC simulation and the theoretical model. The error bars represent the FWHM of the measured 2D angular profile of the guided electron beam.}\label{fig: Fig2}
\end{figure*}
\subsection*{B. Simulations}
We performed 2D3V particle-in-cell (PIC) simulations using the EPOCH code \cite{Arber:2015hc}. Simulations were conducted using p-polarized, 25 fs, 800 nm laser pulses, with both the laser and target parameters closely matching those used in the experiments. Details of the simulation setup and target parameters are provided in the Methods section. The simulation results for both nanostructured and flat targets are presented in Fig. \ref{fig: Fig2}. Panels (a), (b), and (c) are for the nanostructured target, while panels (e), (f), and (g) are for the flat target. Fig. \ref{fig: Fig2}a illustrates the z-component of the magnetic field (Bz) and the trajectories of electrons with energies exceeding 200 keV, shown as black dots, at an angle of incidence (AOI) of 40°, captured 30 fs after the laser-plasma interaction. As shown, on the front side of the structured target, $B_z$ represents the reflected laser and a diffraction mode (discussed later); however, on the rear side, $B_z$ exhibits an azimuthal symmetry, akin to a beam-like current of relativistic electrons. The electron beam deviates from the rear-normal by approximately 30° and consists of a train of sub-fs electron pulses, as evident from their trajectories.

In contrast, for the flat target (Fig. \ref{fig: Fig2}e) a significant portion of the laser light is reflected, leading to low absorption (only 15$\%$, see Fig. S4 of the SI). A fraction of the absorbed energy is coupled to the electrons via JxB heating \cite{Kruer1985}, as evident from the electron trajectories. The electrons are injected in the rear side at every half cycle of the laser pulse, along the laser propagation direction (driven by $\Vec{J}\times\Vec{B}$ force).

Fig. \ref{fig: Fig2}b shows electrons (energy $>$ 300 keV) ejected from the nanopillars toward the rear side at the peak of the pump pulse (15 fs). Notably, we observe a train of sub-femtosecond electron bunches emanating from each nanopillar within the FWHM of the pump laser spot. The longitudinal duration of the electron bunches is as small as 500 attoseconds; however, as they propagate farther, they undergo dispersion and become broadened. The bunch length (p) in the transverse direction depends crucially on the parameters of the nanopillars and the acceleration scheme (as discussed later). It is observed that the period of the electron bunches is approximately 2 fs, which is smaller than the 2.66 fs period of the driving laser pulse (elaborated in the Discussion).

Furthermore, the electron bunches are injected from alternating nanopillars on the rear side (i.e., two injection points at any instant). For comparison, the rear-side electrons (energy $>$ 100 keV) ejected from the flat target are shown in Fig. \ref{fig: Fig2}f. We do not observe such attosecond electron bunches, but a large fraction of electrons are moving with large beam divergence along the normal and laser propagation directions ($\Vec{J}\times\Vec{B}$).

Fig. \ref{fig: Fig2}c demonstrates the rear-side electron angular distribution for five different AOIs of 15°, 25°, 40°, -15°, and -30°, captured 30 fs after the interaction of the peak of the laser pulse with the nanopillar target. The colored triangular shapes depict the direction of the incident laser for each AOI. We observe that a significant fraction of electrons are emitted away from the target normal and J$\times$B directions, which is in excellent agreement with the experimental results of Fig. \ref{fig: Fig3}b. As evident, the electron beams can be guided over a wide angular range by simply changing the AOI of the laser pulse. For comparison, Fig. \ref{fig: Fig2}g shows the rear-side electron angular distribution on the flat target. In the flat target, the electron flux is five times lower, and most of the electrons are emitted along the target normal and a few along the $\Vec{J}\times\Vec{B}$ direction, as expected, again in very good agreement with the experimental results of Fig. \ref{fig: Fig3}c.

Fig. \ref{fig: Fig2}d compares the rear-side electron energy spectra for the nanostructured and flat targets at two AOIs: 15° and 40°. For the nanostructured target at 40°, we observe a high-energy cutoff of approximately 1.2 MeV, which aligns remarkably well with the experimental data (Fig. \ref{fig: Fig3}d). Similarly, the flat target exhibits a measured cutoff energy of 0.5 MeV, closely agreeing with the simulated value of 0.45 MeV. As shown, the cut-off energy also varies with the AOI of the laser pulse. This can be attributed to the sheath field on the target's rear; electrons emitted more toward the normal of the target are decelerated the most compared to electrons moving at larger angles. Fig. \ref{fig: Fig2}h compares the experimentally measured emission angles of the electron beam with the results obtained from PIC simulations and a simple theoretical model (presented in the Discussion section). As evident, the results are in good agreement with both simulation and theory.

\section*{Discussion}
Here, we present a simple theoretical model to explain the principle behind the enhanced acceleration and directionality control of the electron beam. The silicon-nanostructured target used in the study supports two diffraction modes in the far field. One mode is on the front side of the target, as shown in Fig. \ref{fig: Fig2}a, and the other is in the transmission direction on the rear of the target. The angle of diffraction for the two modes can be calculated using the following formula:
\begin{equation}
   \theta_d = \sin^{-1}(\sin\theta_i \pm n\lambda/d)
\end{equation}\label{eq: eq1}
where $\theta_i$ is the AOI of the laser, and d is the period of the nanopillars in the array. However, it is important to note that the solid nanopillars are ionized at the rising edge of the intense femtosecond laser pulse, creating plasma nanopillars. This overdense plasma does not support the diffraction mode on the rear side of the target (as evident in Fig. \ref{fig: Fig2}a). However, surprisingly, we observe that the guided electron beam is directed in the same direction as the diffraction light mode on the rear side, a mode that does not exist for overdense plasma. In Fig. \ref{fig: Fig2}h, we compare the expected diffraction angle for the light mode (without plasma, using eq. 1) with the observed angles of the guided electron beam for different AOIs of the laser pulse. This comparison shows a very good agreement and suggests that the periodic plasma nanopillars can guide the electron beams the same way as the grating does to the light beam, satisfying the same grating equation (eq. 1).

\begin{figure*}[ht]
\centering
\includegraphics[width=1\linewidth]{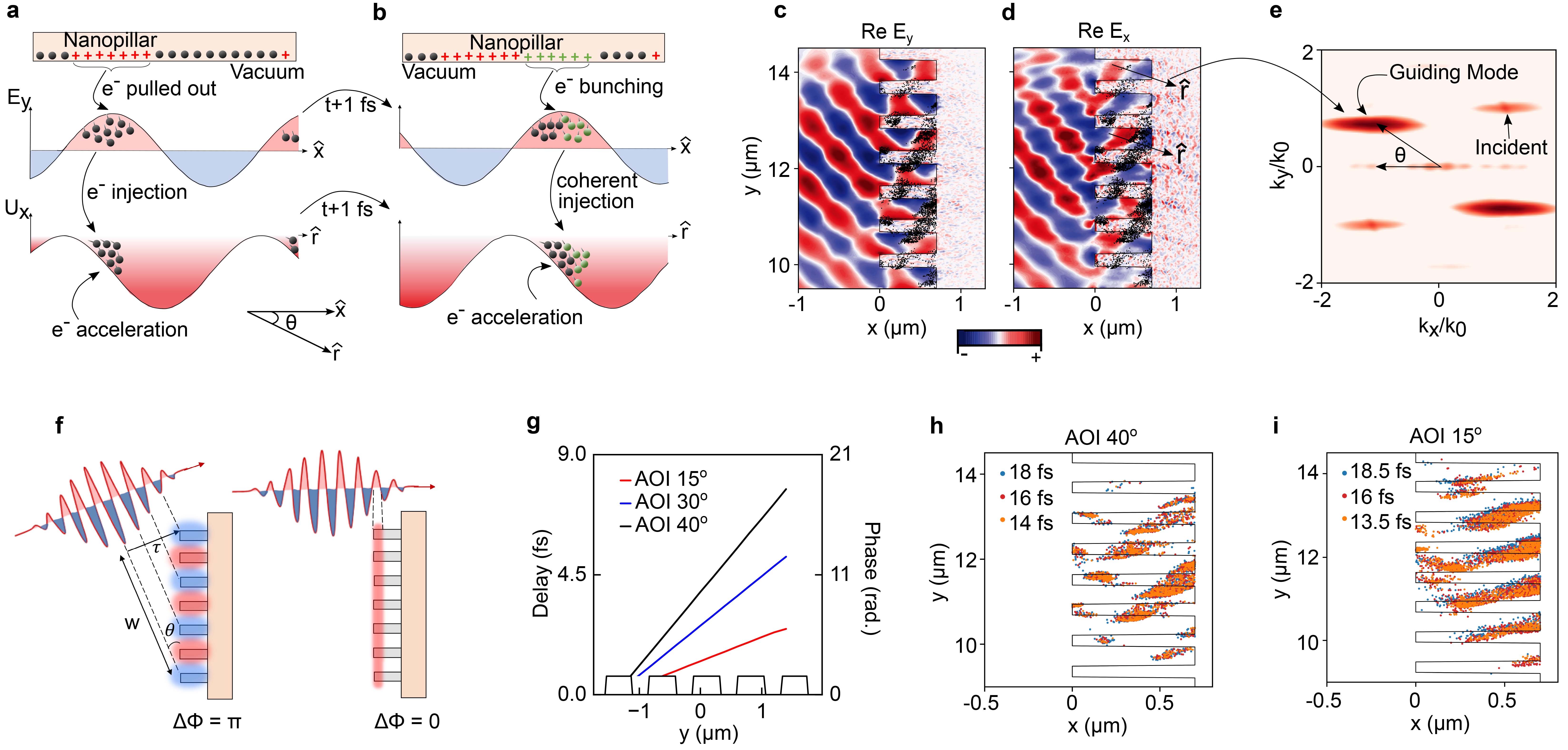}
\caption{\textbf{Guiding and acceleration scheme.} \textbf{a} Schematic showing electron bunches (black circles) pulled from the nanopillar during the positive half-cycle of the $E_y$ field component and injected into the potential energy surface due to the quasi-electrostatic force of the $E_x$ field component along the guiding direction. \textbf{b} Schematic depicting the coherent bunching of electrons from the same nanopillar, where additional electrons (green circles) are pulled by the $E_y$ field component 1 fs later, resulting in the entire bunch gaining energy along the guiding direction. The red and green plus marks represent the positive charge on the nanopillars. \textbf{(c,d)} $E_y$ and $E_x$ components of the field (color plot) near the nanopillar region correspond to the schematic shown in (\textbf{a,b}). The black data points show the electron bunches being pulled outside and accelerated along the guiding direction. \textbf{e} Fourier transform of $E_x$ component of the field in the nanopillar region, specifically showing the dominant guiding mode apart from the incident mode. \textbf{f} Schematic of the relative phase delay induced between neighboring nanopillars as the AOI of the driving laser pulse is changed. \textbf{g} Calculated phase and time delay of laser excitation between nanopillars for three different AOI of $15^{\circ}, 30^{\circ}$, and $40^{\circ}$, respectively. \textbf{(h,i)} Comparison of the dynamics of sub-fs electron bunches in the nanopillar region for two different AOI of $40^{\circ}$ and $15^{\circ}$, respectively.}\label{fig: Fig4}
\end{figure*}

The schematics in Figures \ref{fig: Fig4}a and \ref{fig: Fig4}b depict the electron acceleration mechanism, explaining the enhanced electron energy and the control over the directionality of the electron beam. For a better understanding of the mechanism, we plot the near-field pattern of the total electric field in Figs. \ref{fig: Fig4}c and \ref{fig: Fig4}d. The electric field components $E_y$ (shown in c) and $E_x$ (shown in d) arise from both Mie scattering of the incident light by the nanopillars and the sheath field of the plasma. Notably, the field distribution of both $E_y$ and $E_x$ in between nanopillars is very different from that of the incident laser field. We observe that the phase of $E_y$ between nanopillars propagates along the x direction, while the phase of $E_x$ travels along the $\hat{r}$ direction. The relationship $\hat{r}.\hat{x} = cos(\theta_d)$ holds, where $\theta_d$ is the same as the emission angle of the electron beam. This is also evident from the fast Fourier transform (FFT) of $E_x$ in the nanowire region, shown in Fig. \ref{fig: Fig4}e. As shown, there are two dominant modes: one represents the incident field, and the other is much brighter and along the guiding direction of the electron beam, which is also the same as the angle of diffraction mode at the rear.

 The field of the diffracted laser mode plays a very crucial role in the acceleration and controlling directionality of the electron beam. As depicted in Fig. 4a, during every positive half-cycle of the y-component of the diffracted laser field (Ey), electrons in the upside nanopillar experience a force in the downward direction. As a result, a bunch of electrons are pulled out of the nanopillar, gaining an initial kinetic energy of around 200 keV. Following ionization, this bunch of electrons experience a quasi-electrostatic force along the $\hat{r}$ direction due to the x component of the electric field ($E_x$). The potential landscape, $U_x$ = -$\int E_x \,dr$, experienced by the bunch of electrons due to this force is offset by a phase of 90° to the y-component of the field $E_y$, as shown in the schematic. The electron bunch is injected near the top of the potential hill, where it gains energy as it travels downhill in the potential landscape.

 As the phase of $E_y$ travels forward, more electrons are pulled out of the nanopillar and injected in phase with the earlier bunch in the potential landscape, as shown schematically in Fig. \ref{fig: Fig4}b. However, dephasing inevitably occurs, leading to the loss of some electrons from earlier bunches as the field travels at the speed of light. This way, the electrons are pulled out throughout the whole length of the nanopillar and accelerated together, which are then later injected as sub-fs bunches on the rear side with energy as high as 1.6 MeV. This process repeats with every cycle of the driving laser pulse, resulting in a train of sub-fs electron bunches from each of the nanopillars within the FWHM of the laser spot. In short, even though diffracted laser mode cannot penetrate into the dense plasma target, it is supported by the empty gaps between the nanopillars \cite{luo2018}, where it can guide and further accelerate electron bunches via direct laser acceleration (DLA) 
 \cite{PKS2022,thevenet2016}. Efficient electron acceleration requires phase synchronization of the electron bunch with the diffracted laser mode. However, if the bunch electrons collide with the background plasma between nanopillars, they may become dephased, leading to reduced energy gain (discussed in detail later). Similarly, if the electron bunch is not monoenergetic but has a broad energy spectral width, it leads to spatial dispersion as the bunch travels through the gaps between the nanopillars. Due to this dispersion, electrons within the bunch interact with different phases of the diffracted laser mode, resulting in inefficient electron acceleration. The transverse length (p) of the guided electron bunches significantly exceeds their longitudinal length, as evident in Fig. \ref{fig: Fig2}b. This can be attributed to the differential displacement of electrons originating from different regions of the nanopillar: those ionized and accelerated from the tip are subjected to the force in the emission direction for a longer duration, resulting in greater displacement from the nanopillar in the guiding direction compared to their counterparts ejected from the base.

Figure \ref{fig: Fig4}(f-i) shows how the AOI of the pump pulse can be used to control the phase of excitation. As shown in the schematic of Fig. 4f, when an fs pulse is incident at an angle to the target, the tilt of the pulse front with respect to the target surface causes different spatial regions of the target to be excited at different time delays. This provides a very simple method to control the phase of excitation between the nanopillars of the array. Figure 4g, shows the phase delay experienced by the nanopillars within the FWHM spot of the pump pulse for three different AOIs. We observe that for an AOI of 15°, all the nanopillars are excited with a relative phase difference of less than $\pi$. However, for an AOI of 40°, the neighboring nanopillars of the array are excited with the opposite phase (i.e., $\delta \phi$ = $\pi$), which results in the injection of sub-fs electron bunches from the alternate nanopillar in comparison to an AOI of 15°, as evident from the scatter plot shown in Figs. \ref{fig: Fig4}h and \ref{fig: Fig4}i, respectively. The period of the sub-fs electron bunches emitted from each nanopillar depends on the time interval over which the phase of the electric field $E_y$ oscillates between the nanopillars, following the relation: $t_1$ = $t_2$ cos($\theta$), where $t_1$ and $t_2$ are the periods of the electron bunches and laser pulse, and $\theta$ the AOI of the laser pulse. For an AOI of 40°, the periods of the electron bunch are $t_1$ = 2 fs ($t_2$ = 2.66 fs), and for an AOI of 15°, $t_1$ = 2.5 fs. This is clearly evident from Figs. \ref{fig: Fig4}h and \ref{fig: Fig4}i, as after each period, the electron dynamics repeats both in space and time.

To further demonstrate the validity of the theoretical model, we performed additional PIC simulations with varying target and laser parameters, such as the period and height of the nanopillars. The results are depicted in Fig. 5. Fig. 5a shows the simulated rear-side angular distribution of electrons emitted from the nanopillar target for an AOI of $25^{\circ}$ with varying periods of the nanopillars. Each curve corresponds to a different simulation with a different period of the nanopillars (with a fixed height of 700 nm), as labeled. We can observe a clear shift in the peak emission angle of the electron beam. Fig. 5b compares the mean emission angle of the electron beam obtained from PIC simulations (red points) with the theoretical curve (black solid line) using Equation 1. The mean emission angle was calculated as $\theta_m$ = $\int \theta N(\theta) \,d\theta$ / $\int N(\theta) \,d\theta$, where N($\theta$) is the angular distribution of the guided electron beam. The excellent agreement between the two further validates the physical explanation for the acceleration and guiding mechanism.

\begin{figure}[!t]
\centering
\includegraphics[width=1\linewidth]{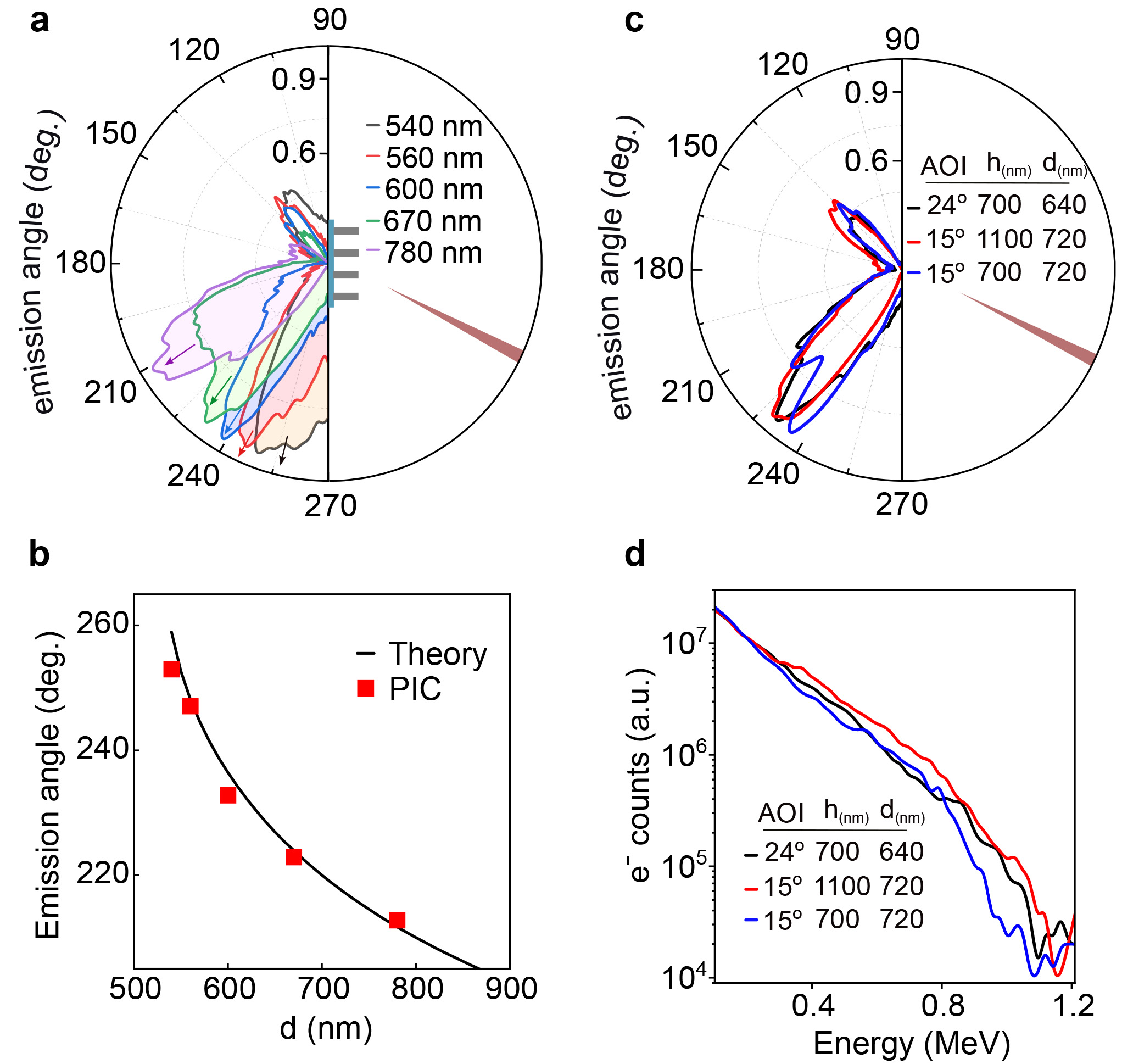}
\caption{\textbf{Effect of varying target parameters on electron dynamics.} \textbf{a} Simulated rear-side angular distribution of electrons for an AOI of 25° with varying periods of the nanopillars, as labeled. \textbf{b} Comparison of the electron emission angles obtained from PIC simulations (red points) with the theoretical model (black line). \textbf{c} Angular distribution and \textbf{d} energy spectrum of rear-side guided electron beams with varying AOI, height (h), and period (d) of the nanopillars, as labeled.}\label{fig: Fig5}
\end{figure}

Figs. 5c and 5d show the angular distribution and the energy spectrum of the guided electron beam with varying AOI, height (h), and period (d) of the nanopillars, as labeled. The red and blue curves in Figs. 5c and 5d correspond to an AOI of 15° with nanopillars having a period of 720 nm but different heights: 1100 nm (red curve) and 700 nm (blue curve). As evident, both the emission angle and the energy spectrum of the electron beam do not change with the change in the height of the nanopillars. The black and blue curves in Figs. 5c and 5d correspond to an AOI of 24 and 15 degrees and a period of 640 and 720 nm, respectively. The angles and the period were chosen using equation 1, in such a way that the direction of the diffracted mode is the same ($\sim$ 238.3°, i.e., 58° away from the normal) for both cases. As observed, electrons in both cases are indeed guided in the same direction, and the emission angle matches very well with the theory. The energy spectrum is also very similar in all these cases.

Another critical factor that can influence the interaction of the laser with the nanostructured target is the pre-plasma induced by the rising edge of the intense femtosecond pulse. This can have a significant effect on electron dynamics, particularly when using lower-contrast laser pulses or high-contrast pulses at intensities exceeding $10^{20}$ W/c$m^2$. We have performed PIC simulation to understand the effect of pre-plasma levels on the emission angle and acceleration mechanisms, and the results are presented in Fig. 6. Fig. 6a shows the 1D density profile of the nanopillars along a line cut along the y-direction, used to initialize the simulation. The pre-plasma density profile is assumed to be exponentially perpendicular to the surfaces of the nanopillars and is defined as $n_e = n_0 exp(-y/L)$, where $n_0$ is the electron number density of the nanopillars and L is the pre-plasma scale length \cite{dulat2022,kluge2018,dulatPOP2024}. Each curve in Fig. 6a corresponds to a different pre-plasma scale length, labeled from 10 to 50 nm. The dashed horizontal line represents the critical density of the laser pulse. Fig. 6b illustrates the fraction of laser energy coupled to the target for different input pre-plasma scale lengths on the nanopillars. As observed, without any pre-plasma, about 67\% of the laser energy is absorbed by the target, and this remains almost unchanged with a pre-plasma scale length of 10 nm. However, there is a decrease in absorption with further increases in scale length, reducing to around 57\% at a pre-plasma level of 40 nm. This decrease in absorption can be attributed to two factors: first, as the scale length increases, the gap between the nanopillars decreases, reducing volumetric heating \cite{Bargsten2017}; second, the vacuum heating efficiency of the laser field on the sidewalls of the nanopillars decreases, as it is most effective only when the quiver amplitude of electrons is larger than the plasma scale lengths \cite{F.Brunel1987}.

\begin{figure}[!t]
\centering
\includegraphics[width=1\linewidth]{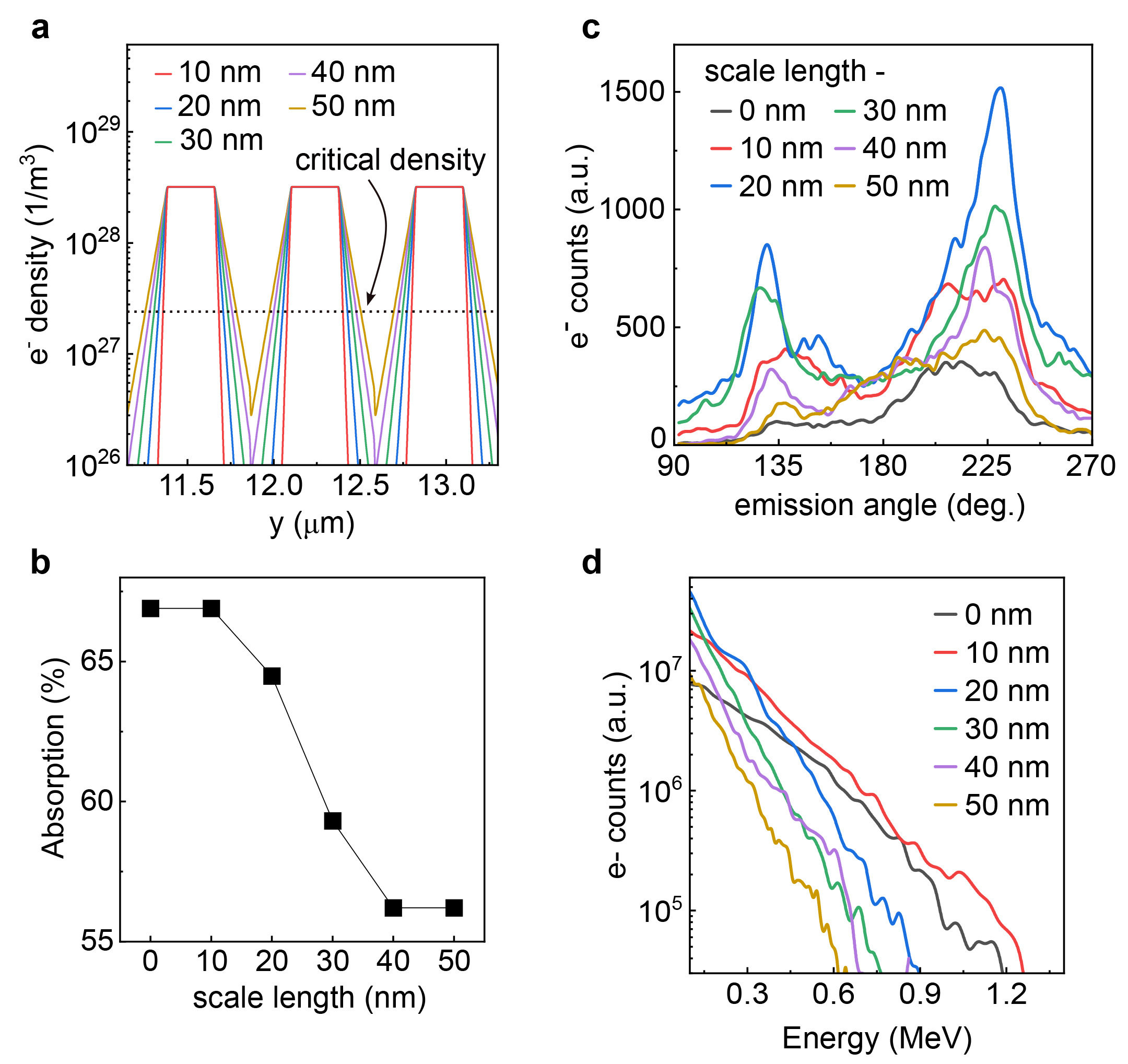}
\caption{\textbf{Pre-plasma effect on electron dynamics.} \textbf{a} Pre-plasma density profile on the nanopillar target used for PIC simulation; each curve corresponds to a different scale length, as labeled. \textbf{b} Change in laser absorption with varying pre-plasma levels on the nanopillar target. \textbf{c} Angular distribution and \textbf{d} energy spectrum of rear-side guided electron beams for varying pre-plasma scale lengths, as labeled.}\label{fig: Fig6}
\end{figure}

In Figs. 6c and 6d, we plot the angular distribution and the energy spectrum of electrons ($>$ 100 keV) at 30 fs after the interaction of the peak of the pulse with the nanopillar target (d = 720 nm, h = 700 nm) at an AOI of 25°. As shown in Fig. 6c, in the absence of pre-plasma, the steered electron beam has a FWHM angular width of around 40° and is centered around 210°. Interestingly, with a pre-plasma scale length of 10 nm, the flux of the steered electron beam doubles, and the peak shifts slightly to 220°. With a pre-plasma scale length of 20 nm, the flux increases by a factor of 5, the angular width reduces to around 22°, and the peak shifts to 230°. Further increases in pre-plasma scale length result in a reduction in flux and an increase in the angular width of the emitted electron beam. Comparing the energy spectrum of the electron beam (shown in Fig. 6d), we observe that the cutoff energy is almost the same for 0 and 10 nm pre-plasma scale lengths. However, with further increases in pre-plasma levels, the cutoff energy decreases, reducing to merely 0.6 MeV for a 50 nm pre-plasma scale length, effectively reduced by a factor of 2. 

The higher flux of electrons emitted for scale lengths of 10 and 20 nm (as shown in Fig. 6c) is primarily due to the fact that, without pre-plasma, only electrons that are pulled out of the nanopillars are injected into the field of the diffraction mode. However, with pre-plasma present, all electrons ahead of the critical surface can directly experience the field of the diffracted laser mode, resulting in the guiding of a larger flux of electrons. However, the acceleration of these electron bunches is not as efficient due to collision-induced dephasing and shorter acceleration lengths (as discussed in the subsequent section). For pre-plasma scale lengths greater than 20 nm, the effective gap between the nanopillars decreases, and the pre-plasma can strongly alter both the amplitude and the spatial profile of the diffracted laser mode, resulting in reduced efficiency in guiding the electron beam. 
The decrease in the cutoff energy of electrons with increasing pre-plasma scale length is primarily due to two reasons. First, as the pre-plasma scale length increases, electrons injected in phase with the diffracted laser modes get dephased more quickly due to collisions with the background plasma (collision-induced dephasing). Second, as the plasma scale length increases, the effective vacuum gap between the nanopillars reduces, limiting the spatial region where electrons can be accelerated before being injected into the overdense plasma. This explains the lower cutoff energies of the electron beam with increasing pre-plasma scale length, as observed in Fig. 6d.

In conclusion, our research bridges the fields of nanophotonics and strong-field plasma physics, facilitating the spatio-temporal control of relativistic electron beams in extreme interactions. We demonstrate, through experiments and simulations, enhanced acceleration and directional emission of electron beams over a wide angular range. Additionally, the crucial role of the phase of excitations on the sub-fs dynamics of electrons in nanopillars is highlighted. We detail the effects of varying target parameters, such as the period and height of nanopillars, on relativistic electron dynamics. Furthermore, the impact of pre-plasma scale length on the laser-nanostructure interaction is explored, revealing that scale length as small as 10s of nm can significantly influence laser coupling, acceleration, divergence, and the guiding mechanism of the electron beam. 

Previously, laser-driven photonic nanostructure-based electron accelerators, known as dielectric laser accelerators \cite{chlouba2023coherent,broaddus2024subrelativistic}, have been demonstrated, which can accelerate electron beams to tens of keV energy over 100s of micrometer distance. Moreover, free electron-light interaction in nanophotonic materials is an active area of research \cite{roques2023free,garcia2021optical,polman2019electron}, as it enables extensive control over the properties of the electron and light beam and also offers nanometer spatial, femtosecond temporal, and millielectron-volt energy resolution in electron-beam spectroscopic studies of nanostructured materials. However, these earlier studies employed laser pulses at much lower intensities, below the ionization threshold of nanostructured materials. In contrast, we have demonstrated coherent control over the dynamics of MeV-energy electron beams via the interaction of relativistic intense laser pulses with plasma nanostructures. This work, therefore, opens exciting possibilities for manipulating laser-driven secondary beams, such as high harmonics and ions, in nanostructured plasmas, advancing the field of laser-plasma interactions and its applications.

\section{Methods}

\subsection*{Target fabrication}
Normally used fabrication techniques like chemical vapor deposition (CVD) or porous anodic alumina (PAA) templates cannot be used to fabricate the perfectly aligned nanopillars we need. These methods tend to produce random nanorod meshes or undesired clustering, making them unsuitable for the controlled experiment. Instead, we fabricate the samples from polycrystalline silicon on a quartz substrate using electron-beam lithography. For sample fabrication, we employ a layer of polycrystalline silicon (poly-Si) created on a four-inch quartz wafer by using low-pressure chemical vapor deposition (LPCVD) in a horizontal tube furnace. The thickness of the Si film is 850 nm, and the thickness of the substrate is 500 micrometers. The pattern is written using a JEOL 9300FS 100kV electron beam lithography (EBL) tool. Polymethyl methacrylate (950K PMMA A4) is used as a positive tone resist. A 10 nm-thick layer of Cr is evaporated on top of the photoresist to prevent charging during the lithography. After the EBL process, wet chemistry removal of the Cr charge dissipation layer, and sequential development, one more 20 nm-thick Cr layer is e-beam evaporated and lifted off to create a hard mask on top of the poly-Si film. The resulting pattern is translated into the surface by means of anisotropic reactive ion etching (RIE) of the silicon layer, which is not masked by chromium. Finally, the Cr mask is removed by wet etching.

\subsection*{Experiment}
The experiment was conducted using the 150 TW, 25 fs, 800 nm laser system at the Tata Institute of Fundamental Research (TIFR), Mumbai. Laser contrast was measured to be $10^{-7}$ at 25 ps before the peak (see Fig. S1b in SI), and the femtosecond temporal profile of the pump pulse is shown in Fig. S1a of the SI. These p-polarized laser pulses were focused on the target using an f/3 off-axis parabolic (OAP) mirror at multiple incidence angles of $15^{\circ}, 25^{\circ}$, and $40^{\circ}$. The measured focal spot of the pump beam was 8 microns, corresponding to a focused peak intensity of 3× $10^{18}$ W/c$m^2$. At this laser intensity, each laser shot damages and ablates the nanostructured sample; therefore, a new sample is used for every shot. As a result, to obtain all the final experimental data presented, approximately 45 to 50 nanostructured samples were used.

An array of silicon elliptical nanopillars arranged in near-wavelength spacing on a 500-micron-thick fused silica substrate was used as the target (shown in Fig. 1). The flat target used for comparison purposes was a 500-micron-thick silicon-fused silica plate. The angular distributions of electrons were measured with imaging plates (IPs) (FUJI Film, BAS-SR 2025) placed in a cylindrical geometry surrounding the target, covering the angular range from 0 to 360 degrees (see Fig. 2a). The IPs were covered with 110-micron-thick aluminum filters to block electrons with energy below 100 keV and to prevent exposure to X-rays, direct lasers, plasma emissions, and ambient light. The angular distributions for each target and angle of incidence (AOI) were obtained with a single laser shot. The energies of fast electrons were measured using electron spectrometers (see Figure 2a) located along two different directions to the target: the front normal (0°) and along the direction of the steered electron beam (210°) in the rear of the target. Each spectrometer has a 0.1 Tesla magnetic field and an IP as a detector. The measurable range of energies in these spectrometers is 0.1–7.0 MeV. Each electron spectrum was obtained by 15 laser shots. The electrons were measured 15 cm away from the target for both angular distribution and energy spectrum measurements. The experiments were performed in a vacuum chamber at a pressure of $10^{-5}$ Torr.\\

\subsection*{PIC Simulation}
We performed a series of 2D3v particle-in-cell (PIC) simulations in Cartesian geometry using the EPOCH code \cite{Arber:2015hc}. The simulation box size is 18$\mu m \times 25\mu m$ with a cell size of 4nm $\times$ 5nm and 16/32 macroparticles per cell. The target used in the experiment was modeled as periodically arranged nanopillars (see Fig. 3) with a period of 720 nm and a height of 700 nm. These parameters are consistent with the target used in the experiment. The plasma consists of electrons with an initial temperature of 100 eV and neutralizing ions with a temperature of 10 eV. To reduce the computational cost and avoid numerical heating, most of the simulation runs were performed with a plasma density of 20 $n_c$. However, we repeated a few runs at 100 $n_c$ to check the precision and convergence of the results. A p-polarized laser pulse with a wavelength of 800 nm irradiates the targets at various incidence angles of 15°, 25°, 40°, -15°, and -30°. The laser pulse is assumed to be Gaussian in both the longitudinal and transverse directions, with a FWHM pulse duration of 25 fs and a beam waist of 5 $\mu$m at the focus. The peak intensity of the focused laser is 3× $10^{18}$ W/c$m^2$, which is the same as the intensity used in the experiment. The laser reaches the target at t = 0 fs. We ran the simulations for 90 fs, 150 fs, and 200 fs, respectively.

\section{Data Availability Statement}
The EPOCH code used for the simulations reported in this paper is publicly available at the following URL: https://github.com/Warwick-Plasma/epoch/releases. Version 4.1.7.16 was used for the simulations. All experimental and simulation data shown in the figures in the article have been deposited in the Figshare data repository (https://doi.org/10.6084/m9.figshare.26044792). All other data is available in the main text or in the SI.

\begin{acknowledgement}

GRK acknowledges partial support from  the  J.C. Bose Fellowship grant JBR/2020/000039 of the Anusandhan National Research Foundation (ANRF), Government of India. Simulations were performed using EPOCH, which was developed as part of the UK EPSRC grants EP/G054950/1, EP/G056803/1, EP/G055165/1 and EP/ M022463/1. This work was supported by Australian Research Council (DE210100679) and by a grant provided by the Australian Academy of Science, on behalf of the Department of Industry, Science and Resources under the Australia-India Strategic Research Fund. The authors acknowledge the TIFR HPC resources used for the simulations reported in this paper.

\end{acknowledgement}


\section{Author Contributions}
GRK conceived the idea and supervised the whole study. S.K. designed and fabricated the nanostructured targets. A.D. performed the experiment with help from S.R., S.D., A.D.L., and Y.M.V..  A.D. analyzed the data. A.D. performed the PIC numerical simulations. The results and interpretation were discussed and finalized by A.D., S.K., and G.R.K..  A.D. wrote the first draft of the manuscript and finalized the manuscript together with S.K. and G.R.K.. All authors contributed to the manuscript.

\begin{suppinfo}
The experimental setup for measuring electron angular distribution, the temporal profile of the pulse, microscopy images of laser shots, measured raw angular distribution data, and laser absorption data.

\end{suppinfo}
\onecolumn
\bibliography{main_ref}

\end{document}